\documentclass[useAMS,usenatbib]{mn2e}
\usepackage{epsfig}
\usepackage{amsbsy}  
\newcommand{\beq}{\begin{equation}}
\newcommand{\eeq}{\end{equation}}
\newcommand{\bea}{\begin{eqnarray}}
\newcommand{\eea}{\end{eqnarray}}
\newcommand{\rde}{\rho_{\rm de}}

\newcommand{\pde}{p_{\rm de}}

\renewcommand{\H}{{\cal H}}
\newcommand{\wt}{{w_{\rm tot}}}
\newcommand{\hcs}{{\hat c}_s^2}
\newcommand{\COSMOMC}{{\sc cosmomc}}
\newcommand{\CAMB}{{\sc camb}}
\newcommand{\CMBFAST}{{\sc cmbfast}}
\newcommand{\Hunit}{~{\rm km}~{\rm s}^{-1} {\rm Mpc}^{-1} }

\newcommand{\deltade}{{\delta_{\rm de}}}

\newcommand{\apj}{Ap.\ J.}
\title[CMB and Dark Energy]
{Large Scale Cosmic Microwave Background Anisotropies and Dark Energy}
\author[J.~Weller and A.M.~Lewis]
{J.~Weller$^1$\thanks{Email: J.Weller@ast.cam.ac.uk} and A.M.~Lewis$^2$\\$^1$Institute of
Astronomy, University of Cambridge, Madingley Road, Cambridge CB3
0HA.\\$^2$CITA, 60 St. George St, Toronto M5S 3H8, ON, Canada}
\date{Accepted ???, Received ???; in original form \today}
\pagerange{\pageref{firstpage}--\pageref{lastpage}}
\pubyear{2003}

\begin{document}
\maketitle

\label{firstpage}

\begin{abstract}
In this note we investigate the effects of perturbations in a dark
energy component with a constant equation of state on large scale
cosmic microwave background anisotropies. 
The inclusion of perturbations increases the large scale power.
We investigate more speculative dark energy models with $w<-1$ and find the opposite
behaviour. Overall the inclusion of perturbations in the dark energy
component increases the degeneracies. We generalise the
parameterization of the dark energy fluctuations to allow for an arbitrary constant sound
speeds and show how constraints from cosmic microwave background
experiments change if this is included.
Combining cosmic microwave background with large scale structure, Hubble
parameter and Supernovae observations we obtain
$w=-1.02\pm 0.16$ (1$\sigma$)  as a constraint on the equation of
state, which is almost independent of the sound speed chosen. With the
presented analysis we find no significant constraint on the constant speed of
sound of the dark energy component.
\end{abstract}
\section{Introduction}
Observations of distant supernovae give strong indications that the
expansion of the universe is accelerating
\citep{Perlmutter:97,Riess:98,Perlmutter98,Riess:01}. This is
consistent with various other evidence, including recent precision
observations of the cosmic microwave background~\citep{Spergel03}. 
These 
observations can in principle be explained by a cosmological constant
term in Einstein's equation of gravity. However, all that is really
required to obtain accelerated expansion of the universe is the
existence of a fluid component which dominates the universe today and
which has a ratio of pressure to energy density of $w\equiv p_{\rm
de}/\rho_{\rm de}<-1/3$. Quintessence models, which assume a
scalar field as the dark energy component \citep{Wetterich:88,Ratra:88,Peebles:88}, differ from a cosmological constant model in that
the equation of state parameter is not necessarily $w=-1$,
and may be evolving. Furthermore a dark energy fluid with
$w\ne -1$ will have perturbations.

In light of the recent cosmic microwave background (CMB) data of
the Wilkinson Microwave Anisotropy Probe (WMAP) \citep{Hinshaw03} we re-investigate the
constraints on a dark energy component with a constant equation of
state and stress the importance of including perturbation in the dark
energy. We note that perturbations have been included in the analysis
of the WMAP team.

If the dark energy is not a cosmological constant, general relativity
predicts that there will be perturbations. Even if dark energy is
expected to be relatively smooth, for a consistent description
of CMB perturbations it is necessary to include perturbations in the
dark energy \citep{Coble:97,Viana:98,Caldwell:98,Ferreira:98}.
We also allow for
models with $w<-1$, as suggested by Caldwell (2002). These models might be realized in
non-minimally coupled scalar field dark energy models
\citep{Amendola:1999qq,Boisseau:00} or k-essence with non-canonical kinetic terms
\citep{Armendariz-Picon:2000dh}. Although the stability of such models is hard
to achieve \citep{Carroll:2003st}, from an observational point of view one
should not rule out the possibility in advance. Recent constraints
from x-ray and type Ia Supernovae observation have constrained the
equation of state to $w=-0.95 \pm 0.30$ \citep{Schuecker:03}.  

As mentioned above, most dark energy scenarios are motivated by
canonical scalar field theories. This leads effectively to
perturbations with a constant speed of sound of the fluctuations with
$\hcs = 1$. However k-essence models allow for an evolving sound
speed~\citep{Armendariz-Picon:2000dh,DeDeo:2003te}. 
We therefore
extend our analysis to models with constant $w$ {\em and} a constant
speed of sound $\hcs$ as a free parameter.

\section{Large scale cosmic microwave anisotropies}
We will concentrate in this analysis on the behaviour of the
temperature anisotropy power spectrum given by the covariance of the
temperature fluctuation expanded in spherical harmonics
\beq
        C_l = 4\pi \int \frac{dk}{k} {\cal P}_\chi |\Delta_l(k,\eta_0)|^2\, .
\eeq
$\Delta_l(k,\eta_0,\mu)$ gives the transfer function for each $\ell$,
${\cal P}_\chi$ is the initial power spectrum
and $\eta_0$ is the conformal time today. 
On large scales the transfer functions are of the form
\beq
\Delta_l(k,\eta_0) = \Delta_l^{\rm LSS}(k) +  \Delta_l^{\rm ISW}(k) \, ,
\eeq
where $\Delta_l^{\rm LSS}(k)$ are the contributions from the last
scattering surface given by the ordinary Sachs-Wolfe effect and the
temperature anisotropy, and  $\Delta_l^{\rm ISW}(k)$ 
is the contribution 
due to the change in the potential $\phi$ along the line of sight and is
called the integrated Sachs-Wolfe (ISW) effect.
The ISW contribution can be written \citep{Sachs:67,Hu95} 
\bea
\Delta_l^{\rm ISW}(k) &=& 2 \int
d\eta\, {\rm
e}^{-\tau(\eta)}
\phi'j_l\left[k(\eta-\eta_0)\right] \nonumber
\eea
where $\tau(\eta)$ is the optical depth due to scattering of the photons along the line of
sight,  $j_l(x)$ are the spherical Bessel functions, and the dash
denotes the derivative with respect to conformal time $\eta$.
The frame-invariant potential $\phi$ can be defined in terms
of the Weyl tensor, and is equivalent to the
 Newtonian potential in the absence of anisotropic stress (see
 \citet{Challinor99} for an overview of the covariant perturbation
 formalism we use here).

The Poisson equation relates the potential to the density
perturbations via
\bea
 k^2 \phi = - 4\pi G  a^2 \overline{\delta\rho}\, ,
\eea
where $\overline{\delta\rho}$ is the total comoving density
perturbation. Thus the source term for the ISW contribution assuming
only matter and dark energy is given by
\bea
 k^2 \phi' =  -4 \pi G \frac{\partial}{\partial \eta}\left[a^2(
\overline{\delta\rho}_m +\overline{\delta\rho}_{\rm de}) \right] \; ,
\label{eqn:ISW}
\eea
where the perturbations are evaluated in the rest frame of the total energy.
The magnitude of the ISW contribution therefore depends on the 
{\em late time evolution} of the total density perturbation.

In general the fractional perturbations
 $\delta_i \equiv \delta\rho_i/\rho_i$ of a non-interacting fluid evolve as
\beq
\delta_i' + 3\H(c_{s,i}^2-w_i)\delta_i + (1+w_i)kv_i = -3(1+w_i)h'\, ,
\label{eqn:pertrho} 
\eeq
where $\H$ is the conformal Hubble parameter, $v_i$ is the velocity,
$w_i\equiv p_i/\rho_i$,  and
$h' = (\delta a/a)'$, where the local scale factor $a$ is defined by
integrating the Hubble expansion. The sound speed $c_s^2$ is
frame-dependent, and defined as $ c_s^2 \equiv \delta p/\delta \rho$.

Neglecting anisotropic stress the potential $\phi$ evolves as
\bea
\phi''+3\H(1+\frac{p'}{\rho'})\phi' + k^2\frac{p'}{\rho'}\phi + \left[
  (1+3\frac{p'}{\rho'})\H^2 + 2\H'\right]\phi \nonumber \\
= 4\pi G a^2(\delta p -
\frac{p'}{\rho'} \delta\rho) \hfill,
\eea
where the RHS is a frame invariant combination.
For a constant total equation of state parameter $\wt$ this becomes
\bea
\phi'' + 3\H(1+\wt)\phi' = 4\pi G a^2 \overline{\delta p}.
\eea
In matter or cosmological constant domination the comoving pressure
perturbation is zero
on scales where the baryon pressure is negligible. In this case
the growing mode is the solution $\phi=\rm{const}$, and there is no contribution
to the ISW effect. However for varying $\wt$, as between matter and dark energy
domination, or when there are dark energy perturbations, the potential
will not be constant.

In general the evolution of the perturbations can be computed
numerically. For a non-interacting fluid with constant $w_i$, defining the frame invariant
quantity $\hat{c}_{s,i}^2$ (the fluid sound speed in the frame
comoving with the fluid) we have the evolution equations
\bea
\delta_i' + 3\H(\hat{c}_{s,i}^2-w_i)(\delta_i +3\H(1+w_i)v_i/k)+\quad\quad \nonumber\\
(1+w_i)kv_i = -3(1+w_i)h' \label{eqn:di1} \\
 v_i' + \H(1-3\hat{c}_{s,i}^2)v_i + kA = k \hat{c}_{s,i}^2
\delta_i/(1+w_i) \, ,
\label{eqn:di2}
\eea
where $A$ is the acceleration ($A=0$ in the $v_m=0$ frame (synchronous gauge), $A=-\Psi$ in the zero shear frame
(Newtonian gauge)). We have assumed zero anisotropic stress, which
is the case for matter and simple dark energy models. Also note that a
varying equation of state factor will lead to extra contributions to
the ISW effect \citep{Corasaniti:03}.

\subsection{Scalar Field Dark Energy}
In order to study the full evolution of the dark energy fluid
including fluctuations we need to specify the
speed of sound and hence its density and pressure perturbations. 
A simple way to achieve this, is by relating the dark energy to a scalar field.
In order to be able to analyse models with an equation of state $w>-1$
as well as $w<-1$ we start with the Lagrangian \citep{Carroll:2003st}
\beq
 {\cal L}_{\rm de} = \pm\frac{1}{2}(\partial_\mu \varphi)^2  - V(\varphi)\; ,
\eeq
where the positive sign in front of the kinetic term corresponds to
$w>-1$ solutions and the negative sign to $w<-1$,
\bea
 \rde = \pm \frac{1}{2}\dot{\varphi}^2 + V\; , \quad\quad
 \pde = \pm \frac{1}{2}\dot{\varphi}^2 - V\, ,
\eea
and dots denote normal time derivatives.
The equations for the perturbations are therefore
\begin{eqnarray} \delta \rde &=& \pm \dot{\varphi}\dot{(\delta\varphi)} + V_{,\varphi}\delta\varphi
\pm A \dot{\varphi}{}^2\\
\delta \pde &=& \pm \dot{\varphi}\dot{(\delta\varphi)} - V_{,\varphi}\delta\varphi
\pm A \dot{\varphi}{}^2
\end{eqnarray}
where $A$ is the acceleration. In the frame in which the scalar field
is unperturbed (the frame comoving with the dark energy, denoted by a
hat),
$\widehat{\delta\varphi}=0$ and so $\hat{c}_s^2 \equiv \widehat{\delta
  p}/\widehat{\delta\rho} = 1$.

If the equation of state $\pde=w\rde$ is
constant, the dark energy density evolves like $\rde = \rho_{{\rm
de},0}\;a^{-3(1+w)}$.  We can then identify
this solution with a scalar field and its potential
\begin{eqnarray}
V(\varphi) & \equiv &\frac{1-w}{2}\rde \; , \label{eqn:V1}\\
 & & \nonumber \\
\dot{\varphi}^2 & \equiv & \pm(1+w)\rde\; .
\label{eqn:phi1}
\end{eqnarray}
Clearly a constant equation of state makes a very unnatural
quintessence model. However a large class of models are expected to be
well described (at least as far as the CMB anisotropy is concerned) by
an effective constant equation of state parameter. In this paper we do
not explicitly consider dark energy models with an evolving equation of state.

In order to analyse the impact of the equation of state parameter of
the dark energy component on the cosmic microwave background
anisotropies we will first look into primary degeneracies originating
from smaller scales in the temperature anisotropy power spectrum. As
discussed in \citet{Melchiorri:2002ux} the main impact is due to the change
in the angular diameter distance toward the last scattering surface. The
small scale CMB anisotropies in a flat universe are mainly sensitive
to the physical cold dark matter and baryon densities and the angular
diameter distance $d_A \propto \int [\Omega_m(1+z)^3+\Omega_{\rm
de}(1+z)^{3(1+w)}]^{-1/2}$. Hence if $w$ is decreasing, we need to increase
$\Omega_{\rm de}$ and for a flat universe decrease $\Omega_{\rm m}$
and therefore increase the Hubble parameter $H_0$ and therefore
decrease $\Omega_b$ in order to obtain the same CMB anisotropy power spectrum.

\begin{figure}
\epsfig{file=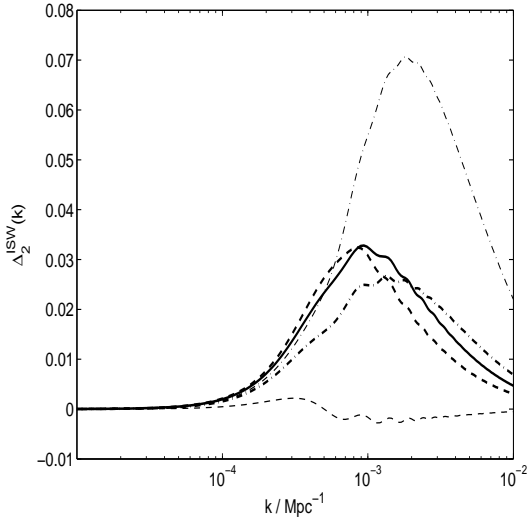,height=7cm,width=7cm}
\caption{The quadrupole ($l=2$) contribution to the integrated
Sachs-Wolfe effect. The solid line is for a $\Lambda$CDM universe, the
dot dashed line for a universe with $w=-2$ and the dashed line for
$w=-0.6$. For the other cosmological parameters see text. The bold
lines are including perturbations in the dark energy component and
the thin lines excluding them.}
\label{fig:ISW2}
\end{figure}

Let us assume that we can by some artificial mechanism suppress the
fluctuations in the dark energy component. Note that in general this
is {\em not} consistent with the equations of general relativity. Only
in the case of a cosmological constant with $w=-1$ we recognise from
Eqn.~\ref{eqn:pertrho} that $\delta\rho_{\rm de} =0$ is a solution. We
implement the equations in the frame comoving
with the dark matter (synchronous gauge), and allow for a changing 
background equation of state but fix the dark energy perturbations to
zero. We compare results from applying this (incorrect) recipe with
those obtained using the full equations consistent with linear general
relativity. In their rest frame the matter perturbations evolve like
\beq
\delta_m''+{\cal H}\delta_m' = 4\pi G a^2 \rho_m
\overline{\delta_m}\,\quad \quad ({\rm forced }\,\delta_{\rm de} = 0) ,
\eeq
which for matter domination ($w=0$) results in $\delta_m \propto a$. If we
gradually decrease $w$ starting from $w=0$, the transition between
matter and dark energy domination 
happens later and later, but more and more rapidly, and with a larger
overall change in the equation of state.
So we expect a  smaller
contribution to the ISW for values of $w$ closer to zero.

In Fig.~\ref{fig:ISW2} we show the quadrupole contribution
$\Delta_2^{\rm ISW}(k)$  to the ISW. The solid line is for a
$\Lambda$CDM universe with $w=-1$, $\Omega_m=0.3$, $\Omega_b=0.05$,
$H_0=65 \Hunit$, the thin dashed line is for $w=-0.6$,
$\Omega_m=0.44$, $\Omega_b=0.073$, $H_0=54 \Hunit$ and the thin dot-dashed for $w=-2$, $\Omega_m=0.17$,
$\Omega_b=0.027$, $H_0=84 \Hunit$. 
For all three models the spectral index is fixed to $n_s=1.0$
and the redshift of instantaneous complete reionization is $z_{\rm re} =17$. 
Without dark energy perturbations we clearly see that for $w=-0.6$
there is only a small contribution to the quadrupole from the ISW,
while there is a large contribution for $w=-2$.

In the case of no dark energy perturbations for $w=-0.6$ there is a smaller ISW
contribution than for a $\Lambda$CDM universe, and subsequently for
$w=-2$ a larger ISW contribution.
\begin{figure}
\epsfig{file=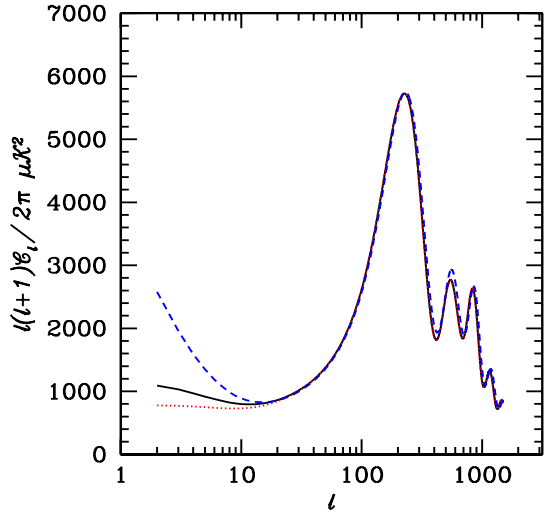,height=7cm,width=7.5cm}
\caption{CMB angular power spectra for different dark energy
models with {\em no} perturbations. The solid line is for a $\Lambda$CDM model, the dotted line
for a model with $w=-0.6$ and dashed line $w=-2.0$. The parameters
$\Omega_c$, $\Omega_b$ and $H_0$ are adjusted to show the
degeneracies as mentioned in the text.}
\label{fig:Clno}
\end{figure}
In Fig.~\ref{fig:Clno} we show the entire temperature anisotropy power
spectrum for the three degenerate models. We can see the
increase in power on large scales by moving from the $w=-0.6$ over the
$w=-1$ ($\Lambda$CDM) to the $w=-2$ model. If these were the true
signatures of dark energy models on large scales we might be hopeful
that by cross correlating large scale CMB anisotropies with x-ray or
radio source power spectra \citep{Boughn:2003yz} one could break the
angular diameter distance degeneracy of the small scale anisotropies. 

The interplay between perturbations in the dark energy and the ISW is
a subtle effect which we will discuss in the section \ref{sec:gen}.
A simple way to understand the opposite behaviour of $w<-1$ models is that
for $w<-1$ the density in the dark energy
component is increasing with an expanding universe, while it is
decreasing in a collapsing universe. Hence the dark energy
perturbations are {\em anti-}correlated with the matter perturbations
as they are sourced. 

The bold lines in
Fig.~\ref{fig:ISW2} correspond to the case which includes
perturbations. Note that for $w=-1$, the perturbations are exactly
zero. We see how the bold dot-dashed line ($w=-2$) is significantly
lowered compared to the thin line, due to the contribution of the
perturbation $\delta\rho_{\rm de}$, while for $w=-0.6$ (dashed line)
the contribution is significantly enhanced.
\begin{figure}
\epsfig{file=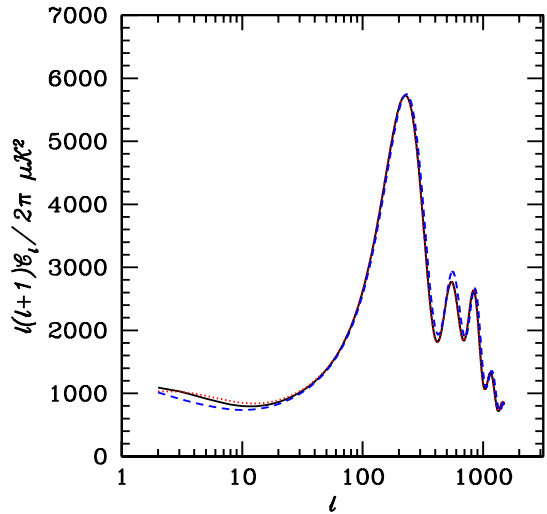,height=7cm,width=7.5cm}
\caption{CMB angular power spectra for them dark energy 
models as in Fig.~\ref{fig:Clno}, but {\em with} dark energy perturbations.}
\label{fig:Clpert}
\end{figure}

In Fig.~\ref{fig:Clpert} we show the CMB temperature anisotropy
spectrum for the three models this time including perturbations. We
clearly see that the large differences obtained on large scales when
we did {\em not} include perturbations in Fig.~\ref{fig:Clno} have
vanished. 
This is because for $w>-1$ the smaller overall change in the
background equation of state is enhanced by the contribution due
to the perturbations in the dark energy
component. For $w<-1$ the large contribution from the different
evolution of the background via the matter perturbations is partially
cancelled by the contribution of the dark energy fluctuation. 
It seems difficult to obtain
information about the nature of dark energy from large scale CMB
information.

\subsection{Generalised Dark Energy Perturbations}\label{sec:gen}
We turn now to the problem of how to describe dark energy perturbations
{\em without} resolving to a scalar field. We should note as a
reminder that we only resolved to a scalar field in order to have a
prescription for calculating the perturbations, where we
assumed the most simple kinetic term $\pm (\partial_u
\varphi)^2$. These models have a speed of sound $\hat{c}_s^2 = 1$.
However we have no idea what the dark energy actually is, so this
assumption may be premature.
For example, in a more generic class of dark energy models, so
called k-essence, the kinetic term does not need to be of such a simple
form \citep{Armendariz-Picon:2000dh} and the sound speed generally differs from one.
In the most general case the speed of sound {\em and} the equation of
state evolve with time, though clearly accounting for this is not feasible in general for parameter
estimation. 
Here we generalise the dark energy parameterisation by introducing a constant sound speed
$\hcs$ as a free parameter. 

\begin{figure}
\epsfig{file=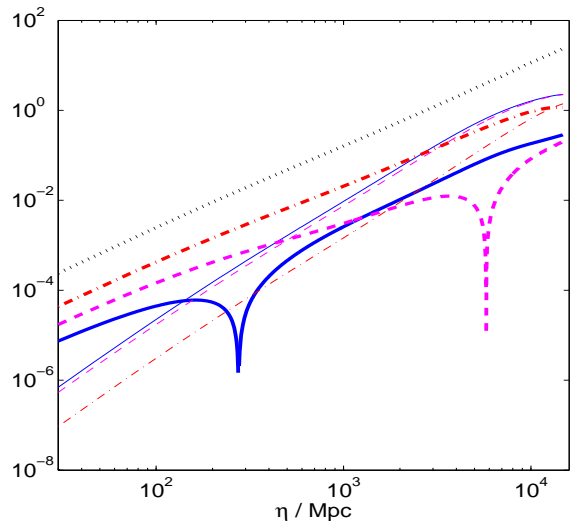,height=7cm,width=7.5cm}
\caption{Evolution of $|\delta_{\rm de}|$ (thick) and $v_{\rm de}$ (thin) in the frame
  comoving with the dark matter perturbation (dotted line), for
  $w=-0.6$ and $\hcs = \{1, 0.7, 0.1\}$ (solid, dashed and dash-dotted lines), and
  $k=10^{-3} {\rm Mpc}^{-1}$. Note that we plot the absolute values of
the fluctuations with amplitude normalized to unit initial curvature perturbation.}
\label{fig:kev}
\end{figure}

If $\deltade$ is initially zero, we see from Eqn.~\ref{eqn:di1} that it
is sourced by the other perturbations if $w\ne -1$ via the time
evolution of the local scale factor, the source term $3(1+w)h'$. An
over density causes a decrease in the local expansion rate
and so $h'<0$. In this case a fluid starts to fall into overdensities
if $w_i>-1$, but starts to fall out if $w_i<-1$. The subsequent evolution depends on the
sound speed, as shown in Fig.~\ref{fig:kev}. Consider the 
frame comoving with the dark matter (where $A=0$). When $k\ll \H$
the term $(1+w_i)kv_i$ can be neglected, then the velocity and
wavenumber only enter via the combination $(1+w_i)v_i/k$. For large
sound speeds the source term for the velocities is large and they are
anti-damped, which leads to an almost $k$-independent evolution
where the dark energy perturbations change sign at early times, and become the
\emph{opposite} 
sign to $\delta_m$. At late times when the dark energy becomes a
significant fraction of the energy density, the total density perturbations
are therefore smaller than without dark energy perturbations, there is
a larger overal change in the potential, and the ISW contribution is increased.
The sign reversal happens later for lower sound speeds as we see in
Fig.~\ref{fig:kev} and for $\hcs \sim 1/3$ the
perturbations never reverse. Thus the contribution to the ISW effect
from the perturbations decreases with the sound speed. For $w<-1$ the
effect is reversed, with the perturbations initially of opposite sign,
and the contribution to the ISW effect increasing as the sound speed
is decreased.

\begin{figure*}
\hrule{\epsfig{file=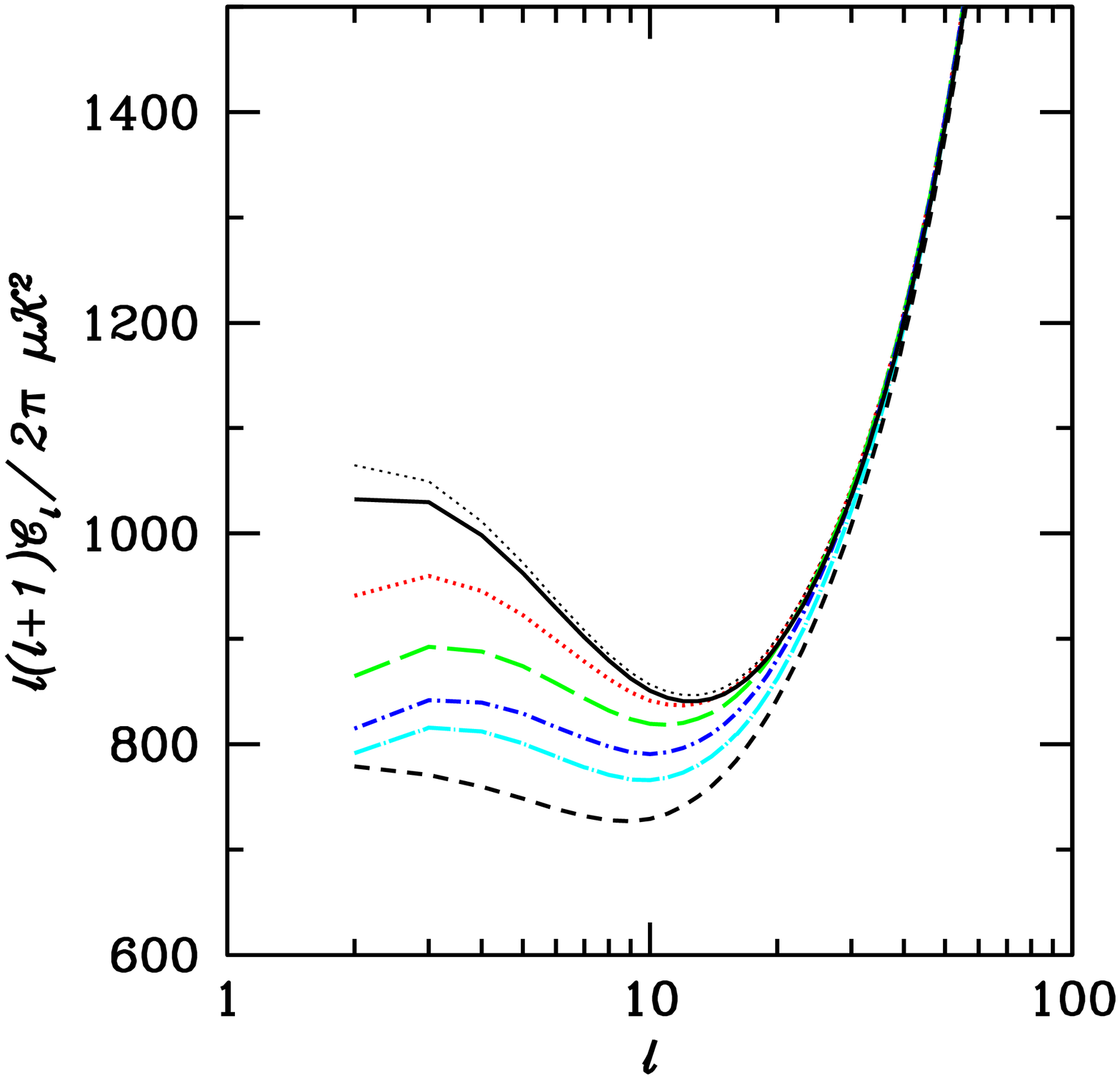,height=7cm,width=7.5cm}\epsfig{file=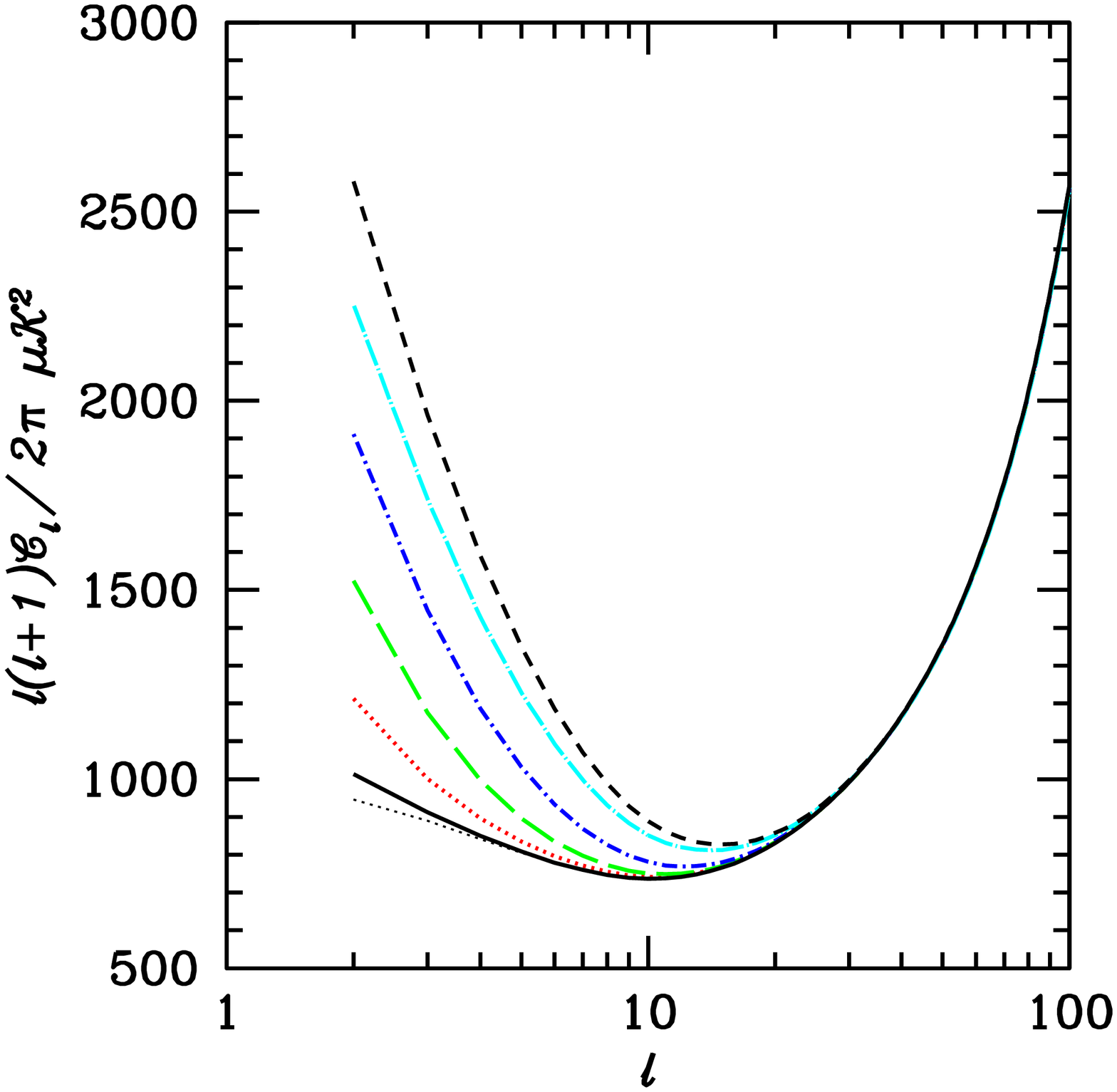,height=7cm,width=7.5cm}}
\caption{On the left the CMB anisotropies for the $w=-0.6$ model. The
top solid line is with perturbations and the low dashed line for
no perturbations. In between the speed of sound is decreasing from top
to down with $c_s^2 = 0.2,0.05,0.01,0.0$. On the right the CMB
anisotropies for the $w=-2.0$ model. The lower solid line is with
perturbations and the top dashed line for no perturbations. In between
the speed of sound is increasing from top to down with $c_s^2 =
0.0,0.01,0.05,0.2$. The thin dotted lines above (for
$w=-0.6$) and below (for $w=-2$) correspond to sound speeds of $c_s^2
= 5.0$. Note that in both cases that $c_s^2 = 1.0$
corresponds to the solid line.} 
\label{fig:cs}
\end{figure*}
In Fig.~\ref{fig:cs} we show how the CMB temperature anisotropies
change on large scales, for different constant $\hcs$. We see that if we
decrease the sound speed gradually from $\hcs=1$ to $\hcs=0$ the ISW
contribution becomes smaller as the dark energy clusters more with the
matter, partly compensating the change in the potential due to the
change in the background equation of state.
Therefore cross - correlating the large scale CMB power spectrum with
direct measures of the potential \citep{Boughn:2003yz} might be an
excellent probe for the sound speed of the dark energy component, if
the equation of state is different from $w=-1$.

\section{Parameter constraints}
In order to stress the importance of the inclusion of dark energy
perturbations we will discuss their impact on the parameter estimation
with CMB data. We included the perturbations into the \CAMB\footnote{http://camb.info} code~\citep{Lewis99}
(based on \CMBFAST~\citep{Seljak96}) and performed a Markov-chain Monte Carlo parameter
analysis using \COSMOMC\footnote{http://cosmologist.info/cosmomc/}~\citep{cosmomc}.
We varied six non-dark energy cosmological
parameters with flat priors: the baryon density $\Omega_b h^2$, the cold dark matter
density $\Omega_{\rm c} h^2$, the ratio of the sound horizon to the
angular diameter distance at last scattering $\theta$, the damping of
the small scale CMB power due to reionization $Z\equiv e^{-2\tau}$ (we
assume $\tau<0.3$), the amplitude of the fluctuations $A_s$ and the
spectral index of the primordial power spectrum $n_s$. In addition we
varied the constant equation
of state parameter of the dark energy component $w$, and where
required the constant sound speed parameter in the range $-3<\log_{10} \hcs<2$. The
Hubble parameter $H_0$ is derived from $\theta$~\citep{Kosowsky02},
and the dark energy density from the requirement that the background
universe is spatially flat. We assume negligible primordial tensor
modes and neutrino mass, and
include priors on the Hubble parameter from the Hubble Key project
\citep{Freedman01}, with $H_0 = (72\pm 8) \Hunit$,
and 
a weak prior $\Omega_b h^2=0.022 \pm 0.002$ (1 $\sigma$) from Big Bang nucleosynthesis \cite{Burles01}.
In addition to the CMB likelihood code provided by WMAP~\citep{Verde03,Hinshaw03,Kogut03} (including the 
temperature-polarization cross-correlation data), we use CBI
\citep{Pearson02} and ACBAR \citep{Kuo02} data for the smaller scales ($\ell>800$). 

\begin{figure}
\epsfig{file=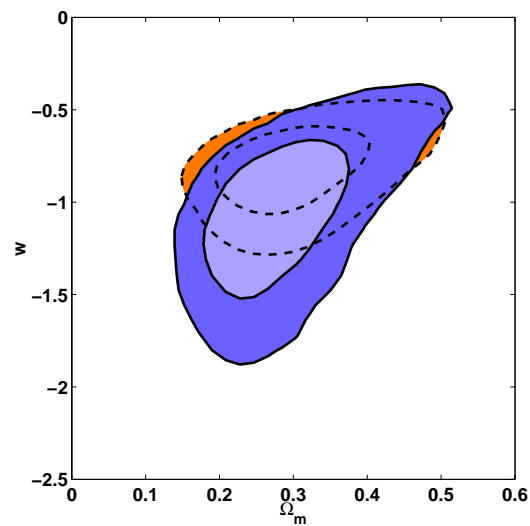,height=7cm,width=7cm}
\caption{Marginalized 68\% and 95\% confidence contours from a combined analysis
of the WMAP, ACBAR and CBI data together with a prior from BBN and
HST, for an (incorrect) smooth dark energy component (dashed lines) and
correctly including perturbations with $\hcs=1$ (solid lines).} 
\label{fig:CMBcont}
\end{figure}

\begin{figure}
\epsfig{file=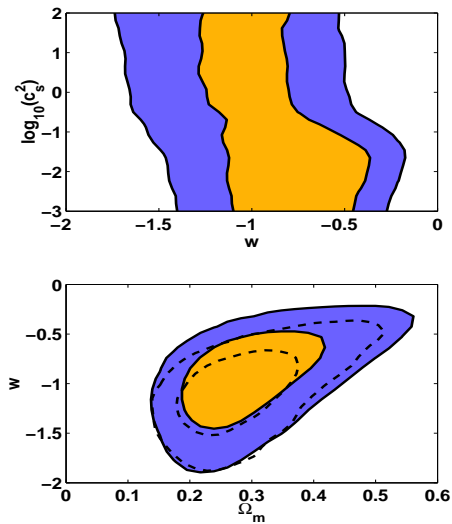,height=7cm,width=6cm}
\caption{Marginalized 68\% and 95\% confidence contours from a combined analysis
of the WMAP, ACBAR and CBI data together with a prior from BBN and
HST, with $\hcs=1$ (dashed) and with $\hcs$ varying (solid).}
\label{fig:CMBcs2}
\end{figure}

In Fig.~\ref{fig:CMBcont} we show the posterior confidence contours in the $\Omega_m-w$ plane. The dashed contours are
from an analysis assuming no perturbations in the dark energy
component, while the solid contours are with perturbations. We clearly
see the different shape of the likelihood contours and how they open
up to more negative values in $w$ if we include perturbations. This is
a direct result of the difference between
Figs.~\ref{fig:Clno} and \ref{fig:Clpert}. Because the large ISW for $w<-1$ is not present
if we include perturbations this part of the parameter space can not
be excluded with CMB data. Furthermore the inclusion of perturbations
leads to more stringent upper bounds on the equation of state
$w$. This is because as we increase the large scale CMB power due to the perturbations
(for $w>-1$), the relatively low quadrupole and octopole disfavour
these models.  In Fig.~\ref{fig:CMBcs2} we show the constraints from
additionally varying a constant sound speed. This slightly favours
values of $w>-1$, where low sound speeds lead to a smaller ISW
contribution at the lowest $\ell$. For $w<-1$ the contours broaden to
include large sound speeds which also give somewhat smaller low multipoles.

\begin{figure}
\epsfig{file=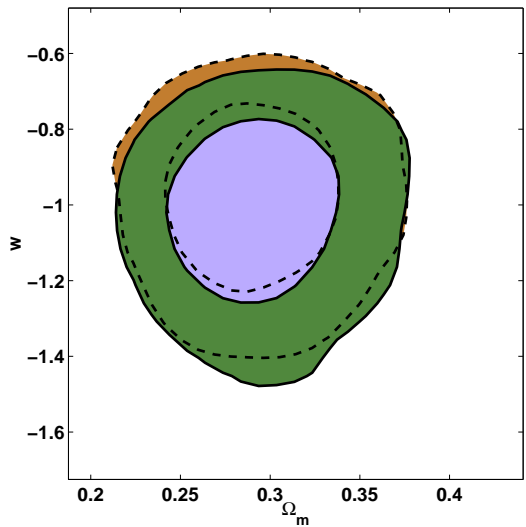,height=7cm,width=7cm}
\epsfig{file=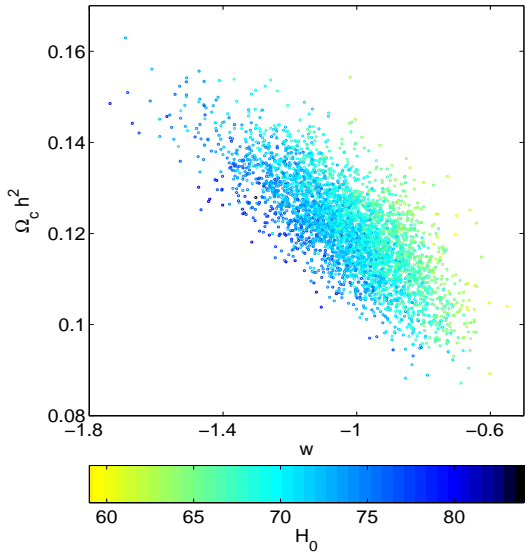,height=8cm,width=7cm}
\caption{Top: 68\% and 95\% contours for a combined analysis of the CMB
data, 2dF, SNe, HST and BBN with $\hcs=1$ (solid) and
marginalizing over $\hcs$ (dashed). Bottom: Samples from the
posterior distribution for $\hcs=1$ with the same data as above. }
\label{fig:CMBall}
\end{figure}
Finally we performed an analysis where we also included the data from the
Supernovae Cosmology Project (SCP) \citep{Perlmutter98} and the two
degree field (2dF) galaxy redshift survey
\citep{Percival01}. The information from the 2dF large scale structure
combined with the prior from the Hubble Key Project constrains the
matter contents, while the Supernovae (SNe) information is
complementary. In Fig.~\ref{fig:CMBall} we show the result of this
combined analysis, with and without marginalizing over a varying sound
speed $\hcs$. The mean value for scalar field models with $\hcs=1$
is $w=-1.02$,  strikingly close
to a cosmological constant, however the 95\% marginalized confidence limit
$-1.37<w<-0.74$ still allows a lot of room for different dark energy
scenarios. Allowing for a different value of the sound speed only slightly shifts the
constraints on $w$ to higher values, with the 95\% result
$-1.32<w<-0.70$. The dominant remaining degeneracies are illustrated
in the scatter plot in Fig.~\ref{fig:CMBall}, where we see how the
constraints depend on the preferred value of the Hubble parameter $H_0$.

\section{Conclusions}
In this note we have re-analysed the constraints on the equation of
state parameter $w$ of dark energy mainly from CMB
observations. We have emphasised the fact that it is essential to
include perturbations in the dark energy component to perform the
analysis. The large scale anisotropies look very different when
perturbations are included and it seems hard to use large scale CMB
information to break the degeneracies.

Furthermore we studied models with an equation of state with
$w<-1$. Our findings are similar to the recently extended version of the WMAP
analysis \citep{Spergel03,Verde03}.
Clearly models with $w<-1$ are under a
lot of pressure for theoretical reasons, since they violate the weak
energy condition and might be unstable. 
However an effective
description of dark energy with non-canonical kinetic terms and a
momentum cut-off might be a valid model for such a scenario
\citep{Armendariz-Picon:2000dh,Carroll:2003st}.  

Finally we found as a posterior mean value for the equation of state
parameter $w=-1.02$, though this conclusion might depend somewhat on our
choice of a  constant equation of state
parameterisation~\citep{Maor:2001ku}. Furthermore we do not find
significant constraints on the value of a constant speed of
sound. We note that in a recent paper \cite{Bean:03} find a $1-\sigma$ detection for a low
sound speed. This is probably due to the fact that they keep
parameters like the the physical matter density fixed. However
cross-correlating the large scale CMB data with large scale structure
measurements could improve these constraints \citep{Boughn:2003yz, Bean:03}.

To conclude a cosmological constant is certainly very consistent with the current
data, however
the 95\% limits on the effective equation of state do not rule out most scalar field dark energy
models. Hence we need better observations to constrain dark energy
models and to be able to distinguish them from a cosmological constant.
While large scale CMB observations are limited by cosmic variance, the
proposed Supernovae Acceleration Probe - SNAP could fulfil this
objective \citep{Weller01}. 
\section*{Acknowledgement} 
We thank S.~Bridle, A.~Challinor, G.~Efstathiou, W.~Hu, M.~Peloso, J.~Ostriker,
P.~Steinhardt and D.~Wands for useful discussions. In the final stages
of this work we became aware that a similar analysis is performed by
R.~Bean and O.~Dor\'e and L.~Boyle, A.~Upadhye and P.~Steinhardt, and we particularly 
thank O.~Dor\'e for useful discussions about that work.
JW is supported by the Leverhulme Trust and a Kings College
Trapnell Fellowship. The parallel computations were done at the UK
National Cosmology Supercomputer Center funded by PPARC, HEFCE and
Silicon Graphics / Cray Research. We further thank the Aspen Center of
Physics, where this work was finalised, for their hospitality.

\label{lastpage}
\end{document}